\documentclass[twocolumn,showpacs,preprintnumbers,amsmath,amssymb]{revtex4}

\usepackage{graphicx}
\usepackage{dcolumn}
\usepackage{bm}

\begin{document}

\title{Strong Scalar QED in Inhomogeneous Electromagnetic Fields}
\author{Sang Pyo Kim}\email{sangkim@kunsan.ac.kr}
\affiliation{Department of Physics, Kunsan National University,
Kunsan 573-701, Korea} \affiliation{Asia
Pacific Center for Theoretical Physics, Pohang 790-784, Korea}

\date{}

\begin{abstract}
Strong QED has attracted attention recently partly because many
astrophysical phenomena have been observed to involve
electromagnetic fields beyond the critical strength for
electron-positron pair production and partly because terrestrial
experiments will generate electromagnetic fields above or near the
critical strength in the near future. In this talk we critically
review QED phenomena involving strong external electromagnetic
fields. Strong QED is characterized by vacuum polarization due to
quantum fluctuations and pair production due to the vacuum
instability. A canonical method is elaborated for pair production at
zero or finite temperature by inhomogeneous electric fields. An
algorithm is advanced to calculate pair production rate for electric
fields acting for finite periods of time or localized in space or
oscillating electric fields. Finally, strong QED is discussed in
astrophysics, in particular, strange stars.\\
\noindent Keywords: Strong QED, Inhomogeneous Electromagnetic field,
Schwinger pair production, Strange star
\end{abstract}
\pacs{12.20.-m, 13.40.-f, 11.10.Wx}

 \maketitle

\section{Introduction}

Recently strong QED (quantum electrodynamics) has attracted much
attention not only from theoretical interest but also from
astrophysical observations and terrestrial experimental tests in the
near future. From a theoretical view point, calculating the full
nonperturbative effective action under the influence of strong
external electromagnetic fields, in particular, inhomogeneous
fields, is still a challenging task (for a recent review and
references, see Ref. \cite{Dunne} and also Ref. \cite{Fradkin}).
From an experimental view point, in the near future electromagnetic
fields from X-ray free electron lasers from LCLS (Linac Coherent
Light Source) at SLAC \cite{slac} and TESLA (TeV Energy
Superconducting Linear Accelerator) at DESY \cite{desy} may attain a
strength almost comparable to the critical value for
electron-positron pair production, which will directly test strong
QED \cite{Ringwald}. Interestingly, astrophysical sources have been
predicted and observed that can have electromagnetic fields greater
than the critical strength. Neutron stars have magnetic fields
ranging from $10^{8}~{\rm G}$ to $10^{15}~{\rm G}$ and more than
one-tenth of them have magnetic fields stronger than $10^{14}~{\rm
G}$, the so-called magnetars (for a review and references, see Ref.
\cite{Wood}), at least one order greater than the critical strength.
Another interesting astrophysical objects with a ultra-strong
electromagnetic field are strange quark stars, hypothetical objects,
which may have electric fields with one or two order higher than the
critical strength \cite{Alcock,Haensel} (see also Ref. \cite{Weber}
for review and references).

Vacuum fluctuations due to a strong external electromagnetic field
contribute nonlinear terms to the classical Maxwell theory and the
electromagnetic theory thus becomes highly nonlinear. Physics in
strong electromagnetic fields drastically differs from the Maxwell
theory \cite{Harding,Marklund}. The cyclotron energy of an electron
in a strong magnetic field can be greater than the rest mass energy
of electron, the equivalent value leading to the critical strength
of magnetic field $B_c = m^2 c^3/e \hbar ~ (4.4 \times 10^{13}~{\rm
G})$. Similarly, in a strong electric field, virtual pairs of
electrons and positrons can gain energy comparable to or greater
than the rest mass energy of electron or positron. The electric
field whose potential energy across the Compton wavelength is the
rest mass energy of electron is the critical value $E_c = m^2 c^3/e
\hbar ~(2.2 \times 10^{15}~{\rm V/cm})$. For magnetic fields greater
than the critical value, nonlinear contributions to the Maxwell term
make the vacuum polarized by quantum fluctuations and the vacuum
polarization causes nonlinear effects such as birefringence
(propagation of photons in the magnetic vacuum), which plays an
important role in the physics of magnetars \cite{Harding}. For
strong electric fields the vacuum decays due to an imaginary part of
the effective action and thus leads to Schwinger pair production
\cite{Schwinger}. Strange stars can emit electron-positron pairs
more efficiently than photons \cite{Usov98,Usov05,Harko06}. On the
other hand, in the standard QED with the minimal interaction
magnetic fields are stable up to $B = 10^{32}~{\rm G}$ due to the
instability from the self-interaction of an electron and up to the
range $B = 10^{51} - 10^{55}~{\rm G}$ due to the instability from
magnetic monopole production at the string or Planck scale
\cite{Duncan}. However, the Pauli interaction may open a window for
pair production by a far weaker inhomogeneous magnetic field and
would have astrophysical applications \cite{Lee-Yoon}.

QED describes the interaction between charged particles and photons.
The success of QED is based on the perturbation theory in the
weak-field limit. However, QED has not been completely understood
yet in the opposite case of strong electromagnetic fields partly
because the full nonperturbative QED action is not known except for
some exactly solved cases \cite{Dunne,Fradkin}. Historically, the
effective action of an electron in a constant electromagnetic field
was obtained by Heisenberg and Euler \cite{Heisenberg-Euler}, and
also by Weisskopf \cite{Weisskopf}. Using the proper time method,
Schwinger found the one-loop effective action for a spin-1/2 fermion
with charge $q$ and mass $m$ in a constant electromagnetic field
\cite{Schwinger}
\begin{eqnarray}
{\cal L}_{\rm eff} &=& - {\cal F} - \frac{1}{8 \pi^2}
\int_0^{\infty}
ds \frac{e^{-m^2s}}{s^3} \nonumber\\
&& \times \Bigl[(qs)^2 {\cal G} \frac{{\rm Re} \cosh (qsX) }{{\rm
Im} \cosh (qsX)} - 1 - \frac{2}{3} (qs)^2 {\cal F} \Bigr].
\end{eqnarray}
where
\begin{eqnarray}
X = [2 ({\cal F} + i {\cal G})]^{1/2} = X_r + i X_i.
\end{eqnarray}
Here, ${\cal F}$ is the negative of the Maxwell term, $- {\cal
L}_{\rm Maxwell}$,
\begin{eqnarray}
{\cal F} = \frac{1}{4} F_{\mu \nu} F^{\mu \nu} = \frac{1}{2} ({\bf
B}^2 - {\bf E}^2),
\end{eqnarray}
and ${\cal G}$ is another Lorentz invariant tensor
\begin{eqnarray}
{\cal G} = \frac{1}{4} F_{\mu \nu} \tilde{F}^{\mu \nu} = {\bf E}
\cdot {\bf B},
\end{eqnarray}
where $\tilde{F}^{\mu \nu} = \epsilon^{\mu \nu \alpha \beta}
F_{\alpha \beta}$ is the dual field tensor. The one-loop effective
action was also obtained in Ref. \cite{Ruffini-Xue}.

The one-loop effective action has two important aspects. First, in
the weak-field limit the nonlinear contribution to the real part
\begin{eqnarray}
{\rm Re}  {\cal L}^{(1)}  = \frac{2}{45 mc^2} \Bigl( \frac{q^2}{4
\pi \hbar} \Bigr)^2 \Bigl(\frac{\hbar}{mc} \Bigr)^3 \Bigl(4 {\cal
F}^2 + 7 {\cal G}^2 \Bigr), \label{real}
\end{eqnarray}
makes the vacuum polarized by quantum fluctuations. In a pure strong
magnetic field the leading term becomes
\begin{eqnarray}
{\cal L}^{(1)} = \frac{(qB)^2}{24 \pi^2} \ln \Bigl(\frac{2 qB}{m^2}
\Bigr).
\end{eqnarray}
The ratio $ {\cal L}^{(1)}/{\cal L}_{\rm Maxwell} = - (q^2/12 \pi^2)
\ln (qB/m^2)$ is the one-loop QED $\beta$-function related with the
renormalization group \cite{Dunne}. Second, in spinor QED in a pure
strong electric field pairs are produced at the rate per unit time
and unit volume
\begin{eqnarray}
w^{\rm fermion} = 2 {\rm Im} {\cal L}^{\rm fermion}_{\rm eff} =
\frac{2}{(2 \pi)^3} \sum_{n = 1}^{\infty} \Bigl(\frac{qE}{n}\Bigr)^2
e^{- \frac{n \pi m^2}{qE}}, \label{fer pr}
\end{eqnarray}
and in scalar QED pairs at the rate
\begin{eqnarray}
w^{\rm boson} = 2 {\rm Im} {\cal L}_{\rm eff}^{\rm boson} =
\frac{2}{(2 \pi)^3} \sum_{n = 1}^{\infty} (-1)^{n +1}
\Bigl(\frac{qE}{n}\Bigr)^2 e^{- \frac{n \pi m^2}{qE}}. \label{bos
pr}
\end{eqnarray}
Here, the factor of $2$ is the spin multiplicity. The scattering
amplitude of the ingoing vacuum to the outgoing vacuum decays
according to Eqs. (\ref{fer pr}) or (\ref{bos pr}), leading to
Schwinger pair production. One way to understand Schwinger pair
production is to compute the imaginary part of the effective action.
Another way is to find the vacuum solution of the field equation and
then calculate the number of pairs produced by the field. The
tunneling interpretation is that charged pairs in the Dirac sea can
tunnel quantum mechanically through the potential barrier lowered by
a uniform electric field. Pair production is an efficient mechanism
for energy extraction from objects with strong electromagnetic
fields.

In this talk, within the framework of canonical quantum field theory
we critically review the Schwinger mechanism at zero or finite
temperature in inhomogeneous electric fields motivated by
terrestrial experiments or astrophysics. Exact solutions of the
Klein-Gordon or Dirac equation minimally coupled to inhomogeneous
electromagnetic fields are known only for a few models. The
Sauter-type electric field that extends for a finite region or lasts
for a finite period is the most well-known model \cite{Nikishov}.
However, one cannot find, in general, solutions for arbitrary
electromagnetic fields, so he or she has to employ some
approximation schemes. It is known that pair production by a
constant electric field that extend over a finite region has a
finite size effect and differs from that by a constant field
\cite{Wang-Wong}. Also the pair production rate by a Sauter-type
electric field obtained by the worldline instanton method depends on
the characteristic scale in a nontrivial way
\cite{Dunne-Schubert,DWGS}. Applying the phase-integral method
\cite{Froman} to find the WKB instanton action for the field
equation with an electric and/or magnetic field in a fixed
direction, the pair production rate is obtained for the Sauter-type
electric field either in space or time with or without a constant
magnetic field \cite{Kim-Page02,Kim-Page06,Kim-Page07}. Further, a
perturbative method is advanced to calculate the WKB instanton
action for pair production by any analytical electric field and is
then applied to strange stars to calculate the production rate of
electron-positron pairs. The thermal effect on the pair production
is also studied \cite{Kim-Lee}.

The organization of this talk is as follows. In Sec. II, we
critically review the Schwinger mechanism and then thermal effects
on pair production. In Secs. III and IV, we apply the WKB instanton
action method to inhomogeneous electric fields that act on for a
finite period or extend for a finite region or oscillate. In Sec.
IV, we apply strong QED to calculate the pair production rate from
strange stars.

\section{Canonical Method for Pair Production}

In real physical systems electric fields are either confined to
finite regions or turned on for finite periods of time. For such
electric fields it is a nontrivial task to calculate the pair
production rate. Instead of applying the proper time method or path
integral method, we employ an approximation scheme such as the WKB
approximation and phase integral in canonical quantum field theory.
For the sake of convenience we consider only scalar QED, but the
formalism here can be directly applied to spinor QED
\cite{Kim-Page06,Kim-Page07}.

\subsection{Schwinger Pair Production}

The Klein-Gordon equation for a charged boson with $q~(q > 0)$ and
$m$ takes the form (in units with $\hbar = c = 1$ and with metric
signature $(+, -, -, -)$)
\begin{eqnarray}
 [\eta^{\mu \nu} (\partial_{\mu} + i q A_{\mu}) (\partial_{\nu}
   + i q A_{\nu}) + m^2] \Phi ({\bf x}, t) = 0. \label{kg fmod}
\end{eqnarray}
Hereafter we further restrict our study to time-dependent electric
fields along the $z$ direction with gauge potentials of the form
$A_z (t) = - E_0 g(t)$ for any analytic function $g(t)$. Then the
Fourier mode, $\Phi ({\bf x}, t) = e^{i {\bf k} \cdot {\bf x}}
\phi_{\bf k}$, satisfies
\begin{eqnarray}
\Bigl[ \frac{\partial^2}{\partial t^2} + m^2 + {\bf k}_{\perp}^2 +
(k_z + q E_0 g(t))^2 \Bigr] \varphi_{\bf k} (t) = 0. \label{kg fmod}
\end{eqnarray}
The solution can be used to quantize the position operators as
\begin{eqnarray}
\hat{\phi}_{\bf k} (t) &=& \varphi_{\bf k} (t) \hat{a}_{\bf k} (t) +
\varphi^*_{\bf k} (t) \hat{b}^{\dagger}_{\bf k} (t),\nonumber\\
\hat{\phi}^*_{\bf k} (t) &=& \varphi_{\bf k} (t) \hat{b}_{\bf k} (t)
+ \varphi^*_{\bf k} (t) \hat{a}^{\dagger}_{\bf k} (t),
\end{eqnarray}
and the momentum operators as
\begin{eqnarray}
\pi_{\bf k} (t) &=& \dot{\varphi}^*_{\bf k} (t) a^{\dagger}_{\bf k}
(t) + \dot{\varphi}_{\bf k} (t) b_{\bf k} (t),\nonumber\\
\pi^*_{\bf k} (t) &=& \dot{\varphi}^*_{\bf k} (t) b^{\dagger}_{\bf
k}(t) + \dot{\varphi}_{\bf k} (t) a_{\bf k} (t).
\end{eqnarray}

On the other hand, in quantum mechanics, Eq. (\ref{kg fmod}) is a
one-dimensional scattering problem with inverted potential. The
positive (asymptotic) solution $\varphi_{{\bf k}, {\rm in}}$ at one
asymptotic region $t = - \infty$ defines the (asymptotic) ingoing
vacuum and another positive (asymptotic) solution $\varphi_{{\bf k},
{\rm out}}$ at the other region $t = \infty$ defines the
(asymptotic) outgoing vacuum. As an incident solution from $t =
\infty$ is partially transmitted over the barrier to $t = -\infty$
and partially reflected by the barrier back to $t = \infty$, the
ingoing solution is related with the outgoing solution as
\begin{eqnarray}
\varphi_{{\bf k}, {\rm in}} = \mu_{\bf k} \varphi_{{\bf k}, {\rm
out}} + \nu_{\bf k} \varphi^*_{{\bf k}, {\rm out}}.
\end{eqnarray}
That is, the ingoing positive frequency solution is mixed both with
the outgoing positive solution and with the outgoing negative
solution, which is the origin of particle production by an external
field \cite{Parker,DeWitt}. As the Wronskian
\begin{eqnarray}
\dot{\varphi}^*_{\bf k} (t) \varphi_{\bf k} (t) - \dot{\varphi}_{\bf
k} (t) \varphi^*_{\bf k} (t) = i,
\end{eqnarray}
is constant, the coefficients satisfy the relation
\begin{eqnarray}
|\mu_{\bf k}|^2 - |\nu_{\bf k}|^2 = 1.
\end{eqnarray}
In fact, the annihilation and creation operators at two asymptotic
regions are related through Bogoliubov transformations
\begin{eqnarray}
\hat{a}_{{\bf k}, {\rm in}} &=& \mu^*_{\bf k} \hat{a}_{{\bf k}, {\rm
out}} -
\nu^*_{\bf k} \hat{b}^{\dagger}_{{\bf k}, {\rm out}}, \nonumber\\
\hat{b}_{{\bf k}, {\rm in}} &=& \mu^*_{\bf k} \hat{b}_{{\bf k}, {\rm
out}} - \nu^*_{\bf k} \hat{a}^{\dagger}_{{\bf k}, {\rm out}}.
\label{in-out}
\end{eqnarray}
The inverse Bogoliubov transformations are
\begin{eqnarray}
\hat{a}_{{\bf k}, {\rm out}} &=& \mu_{\bf k} \hat{a}_{{\bf k}, {\rm
in}} +
\nu^*_{\bf k} \hat{b}^{\dagger}_{{\bf k}, {\rm in}}, \nonumber\\
\hat{b}_{{\bf k}, {\rm out}} &=& \mu_{\bf k} \hat{b}_{{\bf k}, {\rm
in}} + \nu^*_{\bf k} \hat{a}^{\dagger}_{{\bf k}, {\rm in}}.
\label{out-in}
\end{eqnarray}
Therefore, the outgoing vacuum contains the ingoing
particles/antiparticles as \cite{Parker,DeWitt}
\begin{eqnarray}
\langle 0, {\rm out} \vert \sum_{\bf k} \hat{a}^{\dagger}_{{\bf k},
{\rm in}} \hat{a}_{{\bf k}, {\rm in}} \vert 0, {\rm out} \rangle &=&
\sum_{\bf k} |\nu_{\bf
k}|^2, \nonumber\\
\langle 0, {\rm out} \vert \sum_{\bf k} \hat{b}^{\dagger}_{{\bf k},
{\rm in}} \hat{b}_{{\bf k}, {\rm in}} \vert 0, {\rm out} \rangle &=&
\sum_{\bf k} |\nu_{\bf k}|^2,
\end{eqnarray}
and, conversely, the ingoing vacuum evolves into outgoing
particle/antiparticle states as
\begin{eqnarray}
\langle 0, {\rm in} \vert \sum_{\bf k} \hat{a}^{\dagger}_{{\bf k},
{\rm out}} \hat{a}_{{\bf k}, {\rm out}} \vert 0, {\rm in} \rangle
&=& \sum_{\bf k} |\nu_{\bf
k}|^2, \nonumber\\
\langle 0, {\rm in} \vert \sum_{\bf k} \hat{b}^{\dagger}_{{\bf k},
{\rm out}} \hat{b}_{{\bf k}, {\rm out}} \vert 0, {\rm in} \rangle
&=& \sum_{\bf k} |\nu_{\bf k}|^2.
\end{eqnarray}

\subsection{Hamiltonian Approach}

Pair production by time-dependent electric fields can also be
described by the Hamiltonian formalism. The Hamiltonian formalism is
particulary appropriate for studying thermal effects because the
density operator should satisfy the Liouville-von Neumann equation
with respect to the Hamiltonian itself. The complex scalar field has
the Hamiltonian
\begin{eqnarray}
H = \int [d {\bf k}]  \Bigl[\pi_{\bf k}^* \pi_{\bf k} + \omega_{\bf
k}^2 (t) \phi_{\bf k}^* \phi_{\bf k} \Bigr],
\end{eqnarray}
where $[d {\bf k}] = d^3 {\bf k}/(2 \pi)^3$ and
\begin{eqnarray}
\omega^2_{\bf k} (t)  = m^2 + {\bf k}_{\perp}^2 + (k_z + q E_0 g(t)
)^2.
\end{eqnarray}
The field operators are quantized as
\begin{eqnarray}
\hat{\phi} (t, {\bf x}) &=& \int [d {\bf k}] \Bigl[ \varphi_{\bf k}
(t) \hat{a}_{\bf k} (t) + \varphi^*_{\bf k} (t)
\hat{b}^{\dagger}_{\bf k} (t) \Bigr] e^{i {\bf k} \cdot {\bf
x}},\nonumber\\
\hat{\phi}^* (t, {\bf x}) &=& \int [d {\bf k}] \Bigl[ \varphi_{\bf
k} (t) \hat{b}_{\bf k} (t) + \varphi^*_{\bf k} (t)
\hat{a}^{\dagger}_{\bf k} (t) \Bigr] e^{- i {\bf k}. \cdot {\bf x}},
\end{eqnarray}
and the momentum operators as
\begin{eqnarray}
\hat{\pi} (t, {\bf x}) &=& \int [d {\bf k}] \Bigl[
\dot{\varphi}^*_{\bf k} (t) \hat{a}^{\dagger}_{\bf k} (t) +
\dot{\varphi}_{\bf k} (t) \hat{b}_{\bf k} (t) \Bigr]  e^{-i {\bf k}
\cdot {\bf
x}},\nonumber\\
\hat{\pi}^* (t, {\bf x}) &=& \int [d {\bf k}] \Bigl[
\dot{\varphi}^*_{\bf k} (t) \hat{b}^{\dagger}_{\bf k}(t) +
\dot{\varphi}_{\bf k} (t) \hat{a}_{\bf k} (t) \Bigr]  e^{i {\bf k}.
\cdot {\bf x}}.
\end{eqnarray}

In the Hamiltonian approach both quantum states and the density
operator can be found simultaneously using the operators that
satisfy the Liouville-von Neumann equation \cite{Lewis,Kim-Lee00}
\begin{eqnarray}
i \frac{\partial \hat{\rho}_{\bf k} (t)}{\partial t} + [
\hat{\rho}_{\bf k} (t), \hat{H} (t)] = 0. \label{ln eq}
\end{eqnarray}
In fact, there are the time-dependent annihilation and creation
operators satisfying Eq. (\ref{ln eq}) for particles
\begin{eqnarray}
\hat{a}_{\bf k} (t) &=& i \bigl[ \varphi^*_{\bf k} (t)
\hat{\pi}^*_{\bf k} -
\dot{\varphi}^*_{\bf k} (t) \hat{\phi}_{\bf k} \bigr], \nonumber\\
\hat{a}^{\dagger}_{\bf k} (t) &=& - i \bigl[ \varphi_{\bf k} (t)
\hat{\pi}_{\bf k} - \dot{\varphi}_{\bf k} (t) \hat{\phi}^*_{\bf k}
\bigr], \label{pa op}
\end{eqnarray}
and for antiparticles
\begin{eqnarray}
\hat{b}_{\bf k} (t) &=& i \bigl[ \varphi^*_{\bf k} (t)
\hat{\pi}_{\bf k} -
\dot{\varphi}^*_{\bf k} (t) \hat{\phi}^*_{\bf k} \bigr], \nonumber\\
\hat{b}^{\dagger}_{\bf k} (t) &=& - i \bigl[ \varphi_{\bf k} (t)
\hat{\pi}^*_{\bf k} - \dot{\varphi}_{\bf k} (t) \hat{\phi}_{\bf k}
\bigr]. \label{anti op}
\end{eqnarray}
Then the time-dependent vacuum, an exact state of the time-dependent
Schr\"{o}dinger equation, is given by
\begin{eqnarray}
\hat{a}_{\bf k} (t) \vert 0; t \rangle = \hat{b}_{\bf k} (t) \vert
0; t \rangle = 0 \quad ({\rm for~any}~{\bf k}),
\end{eqnarray}
and multi-particle and antiparticle states by
\begin{eqnarray}
\vert n_{{\bf k}_1} \cdots; n_{{\bf k}_2} \cdots; t \rangle =
\frac{\hat{a}^{\dagger n_1}_{{\bf k}_1} (t)}{\sqrt{n_1!}} \cdots
\frac{\hat{b}^{\dagger n_2}_{{\bf k}_2} (t)}{\sqrt{n_2!}} \cdots
\vert 0; t \rangle.
\end{eqnarray}

\subsection{Pair Production at Finite Temperature}

The finite temperature QED effective action was calculated in a
constant magnetic field \cite{Dittrich}, a constant electromagnetic
field \cite{GGF}, and at finite density \cite{EPS}. The Schwinger
proper-time method was used to derive the effective action in a
constant electromagnetic field, which exhibits the Schwinger
mechanism at high temperature \cite{Loewe-Rojas}. However, depending
on the formalism employed to calculate the effective action, pairs
are either produced \cite{Hallin-Liljenberg,GKP} or not produced
\cite{Elmfors-Skagerstam}. The QED effective action from the
imaginary-time formalism has nonzero imaginary part at two-loop
\cite{Gies99}. In this paper we follow the real-time formalism in
Ref. \cite{Kim-Lee} to obtain pair production at finite temperature
at one-loop.

As $\hat{a}_{\bf k} (t), \hat{a}^{\dagger}_{\bf k}$ and
$\hat{b}_{\bf k} (t), \hat{b}^{\dagger}_{\bf k}$ satisfy Eq.
(\ref{ln eq}), the density operator for particles can be found as
\cite{Kim-Lee00}
\begin{eqnarray}
\hat{\rho}_{a_{\bf k}} (t) = \frac{1}{Z_{\bf k}} \exp \Bigl[- \beta
\omega^{\rm in}_{\bf k} \Bigl( \hat{a}^{\dagger}_{\bf k} (t)
\hat{a}_{\bf k} (t) + \frac{1}{2} \Bigr) \Bigr],
\end{eqnarray}
with $\beta = 1/ (kT)$ is the inverse temperature, and for
antiparticles as
\begin{eqnarray}
\hat{\rho}_{b_{\bf k}} (t) = \frac{1}{Z_{\bf k}} \exp \Bigl[- \beta
\omega^{\rm in}_{\bf k} \Bigl( \hat{b}^{\dagger}_{\bf k} (t)
\hat{b}_{\bf k} (t) + \frac{1}{2} \Bigr) \Bigr].
\end{eqnarray}
Then the pair production rate for each spin component from an
initial ensemble by an electric field acting for a finite period of
time is given by \cite{Kim-Lee}
\begin{eqnarray}
n_{\bf k} (E, T) &=& {\rm Tr} \Bigl(\hat{\rho}^{\rm in}_{\bf k}
\hat{N}^{\rm
out}_{a_{\bf k}} \Bigr) - f^{\rm in}_{\bf k} \nonumber\\
&=& |\nu_{\bf k} (E)|^2( 2 f^{\rm in}_{\bf k} (T)  + 1),
\end{eqnarray}
where $f^{\rm in}_{\bf k}$ is the Bose-Einstein distribution
\begin{eqnarray}
f^{\rm in}_{\bf k} (T) = {\rm Tr} \Bigl(\hat{\rho}^{\rm in}_{\bf k}
\hat{N}^{\rm in}_{a_{\bf k}} \Bigr) = \frac{1}{e^{\omega^{\rm
in}_{\bf k}/kT} -1}.
\end{eqnarray}
Here, the Bogoliubov coefficients are
\begin{eqnarray}
\mu_{\bf k} (\infty) &=& i \Bigl(\varphi^*_{\bf k} (\infty)
\dot{\varphi}^{\rm in}_{\bf k}
- \dot{\varphi}^*_{\bf k} (\infty) \varphi^{\rm in}_{\bf k} \Bigr), \nonumber\\
\nu_{\bf k} (\infty) &=& i \Bigl(\varphi^*_{\bf k} (\infty)
\dot{\varphi}^{{\rm in}*}_{\bf k} - \dot{\varphi}^*_{\bf k} (\infty)
\varphi^{{\rm in}*}_{\bf k} \Bigr).
\end{eqnarray}
When there is a uniform magnetic field $B$ in addition to $E(t)$,
each mode of the scalar field obeys the equation
\begin{eqnarray}
\ddot{\varphi}_{nk} (t) + \omega_{nk}^2 (t) \varphi_{nk} (t) = 0,
\end{eqnarray}
where
\begin{eqnarray}
\omega_{nk}^2 = \Bigl(k_z + q E_0 g(t) \Bigr)^2 + qB(2n+1) + m^2.
\end{eqnarray}

\section{Time-Dependent Electric Fields}

Pair production by localized electric fields in time or space
significantly differs from that by the constant electric field due
to a duration or a size effect
\cite{Wang-Wong,Kim-Page02,Dunne-Schubert,Kim-Page06,DWGS}. In this
section we exploit an analytical method to calculate the Schwinger
pair production rate by an electric field acting for a finite period
of time or an oscillating electric field. This case is characterized
by a homogeneous time-dependent electric field $E(t)$ with the
maximum strength $E_0$ and the time scale $T$ defined as
\begin{eqnarray}
T = \frac{1}{2 E_0} \int_{- \infty}^{\infty} E(t) dt.
\end{eqnarray}
The pair production rate is determined by two dimensionless
parameters
\begin{eqnarray}
\epsilon = \frac{m}{qE_0 T}, \quad \delta = \frac{qE_0}{\pi m^2}.
\end{eqnarray}
Pair production is allowed for any $\epsilon$ but is strongly
suppressed for $\epsilon \gg 1$.

The main result of Ref. \cite{Kim-Page07} is that in the weak-field
limit $(E < E_c)$ the mean number of boson pairs for each mode ${\bf
k}$ per unit time and unit volume
\begin{eqnarray}
{\cal N}_{\bf k} = e^{- {\cal S}_{\bf k}}
\end{eqnarray}
is determined by the WKB instanton action of Eq. (\ref{kg fmod})
\begin{eqnarray}
{\cal S}_{\bf k} = i \oint \sqrt{(k_z + q  E_0 g(t) )^2 + m^2 + {\bf
k}^2_{\perp}} dt, \label{inst}
\end{eqnarray}
where the integral is taken outside the contour in the complex plane
of time. In Ref. \cite{Schwinger} the production rate in scalar QED
is defined as twice the imaginary part of the effective action
\begin{eqnarray}
w_{\bf k} = 2 \ln (1 + e^{-{\cal S}_{\bf k}}) = 2 \sum_{n =
1}^{\infty} \frac{(-1)^{n+1}}{n} e^{-n {\cal S}_{\bf k}}.
\label{imag}
\end{eqnarray}
The production rate (\ref{imag}) is even valid for very strong
electric fields $(E \gg E_c)$ or $({\cal S}_{\bf k} \ll 1)$.

A few comments are in order. Taking into account spin statistics,
the mean number of boson pairs with spin $s$ is given by
\cite{Kim-Page05}
\begin{eqnarray}
{\cal N}^{\rm boson}_{\bf k} = e^{w^{\rm boson}_{\bf k}} - 1 =  (1 +
e^{-{\cal S}^{\rm boson}_{\bf k}})^{2s +1} -1, \label{boson mean
pair}
\end{eqnarray}
while the mean number of fermion pairs with spin $s$
\begin{eqnarray}
{\cal N}^{\rm fermion}_{\bf k} = 1 - e^{- w^{\rm fermion}_{\bf k}} =
1 - (1 - e^{- {\cal S}^{\rm fermion}_{\bf k}})^{2s+1},
\label{fermion mean pair}
\end{eqnarray}
where $w^{\rm boson/fermion}_{\bf k}$ is twice the imaginary part of
the effective action for bosons and fermions, respectively, and $2s
+1$ is the spin multiplicity. The mean numbers without the spin
multiplicity is also obtained in Ref. \cite{Gavrilov}. The mean
numbers (\ref{boson mean pair}) and (\ref{fermion mean pair}) hold
even for very strong electric fields. The first term  in Eq.
(\ref{boson mean pair}) is the amplification factor for boson
production. In the weak-field limit $({\cal S}_{\bf k} > 1)$, the
mean number of boson or fermion pairs is approximately given by
\begin{eqnarray}
{\cal N}^{\rm boson/fermion}_{\bf k} \approx (2s+1) e^{- {\cal
S}^{\rm boson/fermion}_{\bf k}}. \label{fermion pair}
\end{eqnarray}
Another interesting point is the relation between the WKB instanton
actions for bosons and fermions. For a Sauter electric field it is
shown in Ref. \cite{Kim-Page07} that the WKB instanton action for
fermions together with the next-to-leading order is nothing but the
WKB instanton action for bosons. Such a relation plausibly holds for
any inhomogeneous electric field since the one-loop effective action
for fermions
\begin{eqnarray}
S^{\rm fermion} = -i \ln \det (i {\bf D} -m), \quad {\bf D} =
\gamma_{\mu}
\partial_{\mu} + i q A_{\mu},
\end{eqnarray}
has the imaginary part, which is approximately for bosons
\begin{eqnarray}
{\rm Im} (S^{\rm fermion}) &=& -\frac{i}{2} [\ln \det (i {\bf D} -m)
+ \ln \det (i
{\bf D} -m)^{*}] \nonumber\\
&\approx& -\frac{i}{2} \ln \det(D^2 +m^2).
\end{eqnarray}
Therefore, the WKB instanton action works for both scalar and spinor
QED.

Now we develop an algorithm to compute the instanton action
systematically. Introducing another variable
\begin{eqnarray}
\zeta = g (t),
\end{eqnarray}
we rewrite Eq. (\ref{inst}) as
\begin{eqnarray}
{\cal S}_{\bf k} = i (qE_0) \oint \sqrt{\Bigl(1 + \frac{k_z}{qE_0
\zeta} \Bigr)^2 + \frac{m^2 + {\bf k}^2_{\perp}}{(qE_0 \zeta)^2}}
\frac{ \zeta  d \zeta}{g' (\zeta)}, \label{inst2}
\end{eqnarray}
and expand the square root in an inverse power series
\begin{eqnarray}
\sqrt{\Bigl(1 + \frac{k_z}{qE_0 \zeta} \Bigr)^2 + \frac{m^2 + {\bf
k}^2_{\perp}}{(qE_0 \zeta)^2}} = \sum_{n = 0}^{\infty} \frac{{\cal
C}_n}{\zeta^n},
\end{eqnarray}
and the function $1/g'(\zeta)$ in a power series
\begin{eqnarray}
\frac{1}{g' (\zeta)} = \sum_{n = 0}^{\infty} {\cal D}_n \zeta^n.
\end{eqnarray}
Then the sum of negative residues of simple poles leads to the WKB
instanton action
\begin{eqnarray}
{\cal S}_{\bf k} = 2 \pi (qE_0) \sum_{n = 0}^{\infty} {\cal C}_{n+2}
{\cal D}_n.
\end{eqnarray}
The first few terms of ${\cal C}_n$ are
\begin{eqnarray}
{\cal C}_0 &=& 1, \quad {\cal C}_1 = \alpha_1, \quad
{\cal C}_2 = \frac{\alpha_2}{2}, \nonumber\\
{\cal C}_3 &=& - \frac{\alpha_1 \alpha_2}{2}, \quad {\cal C}_4 =
\frac{\alpha_1^2 \alpha_2}{2} - \frac{\alpha_2^2}{8},
\end{eqnarray}
where
\begin{eqnarray}
\alpha_1 = \frac{k_z}{qE_0}, \quad \alpha_2 = \frac{m^2 + {\bf
k}^2_{\perp}}{(qE_0)^2}.
\end{eqnarray}
The coefficients ${\cal D}_n$ are determined by the profile of
$g(t)$. For specific models, we consider a Sauter-type electric
field $E (t) = E_0~{\rm sech}^2 (t/T)$ and an oscillating electric
field $E(t) = E_0~\cos (t/T)$.

\subsection{Sauter-Type Electric Field}

The gauge potential for $E(t) = E_0 ~{\rm sech}^2 (t/T)$ in the $z$
direction is given by the Sauter potential
\begin{eqnarray}
A_z (t) = - E_0 T \tanh \Bigl(\frac{t}{T} \Bigr).
\end{eqnarray}
With the change of variable $\zeta = g (t) = T \tanh (t/T)$, we have
the power series
\begin{eqnarray}
\frac{1}{g' (t)} = \frac{1}{1 - \frac{\zeta^2}{T^2}} = \sum_{n =
0}^{\infty} \frac{\zeta^{2n}}{T^{2n}},
\end{eqnarray}
and find $ {\cal D}_{2n} = 1/T^{2n}$. This means that the WKB
instanton action is
\begin{eqnarray}
{\cal S}_{\bf k} = 2 \pi (qE_0 T^2) \sum_{n = 0}^{\infty}
\frac{{\cal C}_{2n+2}}{T^{2n+2}}. \label{sauter sum}
\end{eqnarray}
In terms of the scaled variables and parameters
\begin{eqnarray}
\lambda \equiv \frac{k_z}{qE_0 T}, \quad \kappa \equiv
\frac{k_{\perp}}{m}, \quad Z = 2 \pi q E_0 T^2 = \frac{2}{\delta
\epsilon^2},
\end{eqnarray}
the leading terms of the WKB instanton action are
\begin{eqnarray}
{\cal S}_{\bf k} = Z \epsilon^2  (1+ \kappa^2) \Bigl[ \frac{1}{2} +
\frac{\lambda^2}{2} - \frac{\epsilon^2 (1 + \kappa^2)}{8} + \cdots
\Bigr].
\end{eqnarray}
In fact, the sum (\ref{sauter sum}) can be done exactly as
\cite{Kim-Page07}
\begin{eqnarray}
{\cal S}_{{\bf k}_{\perp}} &=& \frac{Z}{2} \Bigl[ \sqrt{(1+
\lambda)^2 + \epsilon^2 (1 + \kappa^2) } \nonumber\\&& + \sqrt{(1 -
\lambda)^2 + \epsilon^2 (1 + \kappa^2)} - 2 \Bigr]. \label{sp ins}
\end{eqnarray}
The instanton method \cite{Kim-Page07} gives much closer result to
the exact one \cite{Nikishov} than the worldline instantons
\cite{DWGS}.

\subsection{Oscillating Electric Field}

The oscillating electric field $E (t) = E_0~ \cos (t/T)$ has many
physical applications such as laser fields
\cite{Brezin-Itzykson,Popov72,Popov02}.  The electric field from
oscillating plasma due to pair production is approximately given by
$E(t) = E \cos (t/T)$, where $E$ varies slowly during the
oscillation period \cite{KESCM,Cooper-Mottola,RVX03,RVX07}. The
gauge potential $\zeta = g(t) = T \sin(t/T)$ leads to the expansion
\begin{eqnarray}
\frac{1}{g'(t)} = \frac{1}{\sqrt{1 - \frac{\zeta^2}{T2}}} = 1 +
\frac{\zeta^2}{2T^2} + \frac{3 \zeta^4}{8 T^4} + \frac{5 \zeta^6}{16
T^6} + \cdots.
\end{eqnarray}
Repeating the procedure in Sec. III A, we obtain the leading terms
of the WKB instanton action
\begin{eqnarray}
{\cal S}_{\bf k} = Z \epsilon^2  (1+ \kappa^2) \Bigl[ \frac{1}{2} +
\frac{\lambda^2}{4} - \frac{\epsilon^2 (1 + \kappa^2)}{16} + \cdots
\Bigr].
\end{eqnarray}

\section{Inhomogeneous Electric Field}

For an inhomogeneous electric field localized in the $z$ direction,
we may choose a Coulomb gauge $A_0 (z) = - E_0 h(z)$, which leads to
$E(z) = E_0 h'(z)$. Then the Klein-Gordon equation has the Fourier
mode solution, $\Phi ({\bf x}, t) = e^{i {\bf k}_{\perp} \cdot {\bf
x}_{\perp} - i \omega t} \phi_{{\bf k}_{\perp}}$, given by
\begin{eqnarray}
\Bigl[- \frac{\partial^2}{\partial z^2} + m^2 + {\bf k}_{\perp}^2 -
(\omega + q E_0 g(z))^2 \Bigr] \varphi_{{\bf k}_{\perp}} (z) = 0.
\label{kg hmod}
\end{eqnarray}
The characteristic length scale $L$ is defined as
\begin{eqnarray}
L = \frac{1}{2 E_0} \int_{- \infty}^{\infty} E(z) dz.
\end{eqnarray}
As for time-dependent electric fields, two dimensionless parameters
\begin{eqnarray}
\bar{\epsilon} = \frac{m}{qE_0 L}, \quad \delta = \frac{qE_0}{\pi
m^2},
\end{eqnarray}
determine the pair production rate. Remarkably the pair production
rate is again given by the WKB instanton action \cite{Kim-Page07}
\begin{eqnarray}
{\cal S}_{\bf k} = - i \oint \sqrt{(\omega + q  E_0 h(z) )^2 - (m^2
+ {\bf k}^2_{\perp})} dz, \label{inst sp}
\end{eqnarray}
where the integral is taken outside the contour in the complex plane
of space. We again introduce the variable
\begin{eqnarray}
\zeta = h (z),
\end{eqnarray}
and rewrite Eq. (\ref{inst sp}) as
\begin{eqnarray}
{\cal S}_{\bf k} = - i (qE_0) \oint \sqrt{\Bigl(1 +
\frac{\omega}{qE_0 \zeta} \Bigr)^2 - \frac{m^2 + {\bf
k}^2_{\perp}}{(qE_0 \zeta)^2}} \frac{ \zeta d \zeta}{g' (\zeta)}.
\label{inst sp2}
\end{eqnarray}
The difference from the case of time-dependent electric fields is
that the overall sign changes, and $\alpha_1$ and $\alpha_2$ are now
replaced by $\bar{\alpha}_1  = \omega/ (qE_0)$ and $\bar{\alpha}_2 =
- \alpha_2$.

For specific models, we consider first a Sauter-type electric field
$E (z) = E_0~{\rm sech}^2 (z/L)$ and then the electric field from
strange stars in the next section. The Coulomb gauge is given by the
Sauter potential
\begin{eqnarray}
A_0 (z) = - E_0 L \tanh \Bigl(\frac{z}{L} \Bigr).
\end{eqnarray}
With $\zeta = L \tanh (z/L)$ and the replacement of $\alpha_1$ by
$\bar{\alpha}_1$ and $\alpha_2$ by $\bar{\alpha}_{2} = -\alpha_2$,
the leading terms of the WKB instanton action are
\begin{eqnarray}
{\cal S}_{\bf k} = \bar{Z} \bar{\epsilon}^2  (1+ \kappa^2) \Bigl[
\frac{1}{2} + \frac{\bar{\lambda}^2}{2} + \frac{\bar{\epsilon}^2 (1
+ \kappa^2)}{8} + \cdots \Bigr].
\end{eqnarray}
where
\begin{eqnarray}
\bar{\lambda} \equiv \frac{\omega}{qE_0 L}, \quad \kappa \equiv
\frac{k_{\perp}}{m}, \quad \bar{Z} = 2 \pi q E_0 L^2 =
\frac{2}{\delta \bar{\epsilon}^2}.
\end{eqnarray}
The exact sum of (\ref{sauter sum}) is known \cite{Kim-Page07}
\begin{eqnarray}
{\cal S}_{{\bf k}_{\perp}} &=& \frac{\bar{Z}}{2} \Bigl[2 - \sqrt{(1+
\bar{\lambda})^2 + \bar{\epsilon}^2 (1 + \kappa^2) } \nonumber\\&&
-\sqrt{(1 - \bar{\lambda})^2 + \bar{\epsilon}^2 (1 + \kappa^2)}
\Bigr]. \label{sp ins}
\end{eqnarray}

\section{Pair Production from Strange Quark Stars}

A source of the most strong electric fields beyond the critical
strength is strange quark stars. A quark star, a hypothetical
astrophysical object, could be formed from a hadron-quark phase
transition at high densities and/or temperatures
\cite{Itoh,Bodmer,Witten}. Chemical equilibrium of $u, d$ and $s$
quark and the charge neutrality of strange stars requires a net
amount of electrons that are free to move the surface but bounded by
the electric attraction from the positive core, thus forming an
electrosphere of several hundred fermis. Then, the electrosphere of
strange stars can generate an extremely strong electric field as
strong as $5 \times 10^{17}~{\rm V/cm}$, two order greater than the
critical strength, and leads to efficient production of
electron-positron pairs \cite{Usov98,Usov05,Harko06}.

The static potential from the Thomas-Fermi model of the electron
distribution at temperature $T$ \cite{Alcock}
\begin{eqnarray}
\frac{d^2 V}{dz^2} &=& \frac{4 \alpha}{3 \pi} \Bigl[(V^3 - V^3_q) +
\pi^2 T^2 (V - V_q ) \Bigr], \quad (z \leq 0), \nonumber\\ \frac{d^2
V}{dz^2} &=& \frac{4 \alpha}{3 \pi} \Bigl[V^3 + \pi^2 T^2 V \Bigr],
\quad (z \geq 0),
\end{eqnarray}
where $z$ is the coordinate normal to the quark surface, $\alpha$ is
the fine structure constant, and $V_q/ 3 \pi^2$ is the quark charge
density inside the quark matter. The boundary condition is $V (-
\infty) = V_q$, $V (\infty) = 0$, and $V(0) = 3 V_q/4$. The Coulomb
gauge potential is found \cite{Harko06}
\begin{eqnarray}
A_0(z) = \frac{\sqrt{2} \pi T}{\sinh \Bigl[ 2 \sqrt{\frac{\alpha
\pi}{3}} T (z +z_0) \Bigr]},
\end{eqnarray}
and the electric field is
\begin{eqnarray}
E(z, T) = \sqrt{\frac{8 \pi^3}{3}} T^2 \frac{\cosh [2
\sqrt{\frac{\alpha \pi}{3}} T(z +z_0)]}{\sinh^2 [2
\sqrt{\frac{\alpha \pi}{3}} T(z +z_0)]},
\end{eqnarray}
whose characteristic scales are
\begin{eqnarray}
E_0 = \sqrt{\frac{8 \pi^3 \alpha}{3}} T^2, \quad L =
\sqrt{\frac{3}{\alpha \pi}} \frac{1}{2T}.
\end{eqnarray}
The higher (lower) the temperature is, the greater (smaller) is the
maximum strength and the narrower (wider) is the width of the
electric field.

With
\begin{eqnarray}
\zeta = - \frac{L}{\sinh \Bigl[\frac{z +z_0}{L} \Bigr]},
\end{eqnarray}
the WKB instanton action becomes
\begin{eqnarray}
{\cal S}_{\bf k} &=& - i (qE_0 L) \oint \sqrt{\Bigl(1 +
\frac{\omega}{qE_0 \zeta} \Bigr)^2 - \frac{m^2 + {\bf
k}^2_{\perp}}{(qE_0 \zeta)^2}} \nonumber\\ && \times \frac{d
\zeta}{\sqrt{1 + \frac{\zeta^2}{L^2}}}. \label{inst sp2}
\end{eqnarray}
Thus, in terms of the scaled variables and parameters in Sec. IV,
the leading terms of the WKB instanton action are
\begin{eqnarray}
{\cal S}_{\bf k} &=& 2 \pi (q E_0 L^2) \Bigl[ \frac{\omega}{qE_0 L}
- \frac{1}{4} \frac{\omega}{q E_0 L} \frac{m^2 + {\bf
k}_{\perp}^2}{(qE_0L)^2} + \cdots \Bigr] \nonumber\\
&=& \bar{Z} \bar{\lambda} \Bigl[ 1 - \frac{\bar{\epsilon}^2 (1 +
\kappa^2)}{4} + \cdots \Bigr].
\end{eqnarray}
Then, the mean number (\ref{fermion mean pair}) of electron-positron
pairs per unit time and volume
\begin{eqnarray}
{\cal N}^{\rm fermion}_{\bf k} = 2 e^{- {\cal S}_{\bf k}} - e^{- 2
{\cal S}_{\bf k}}
\end{eqnarray}
is the spectrum of emitted pairs. As $ \bar{Z} \bar{\lambda} =
\sqrt{3 \pi/ \alpha} \times (\omega/T)$, hot strange stars produce
more pairs of electrons and positrons than cold ones, confirming the
numerical result of Ref. \cite{Harko06}.

\section{Conclusion}

In this talk, we critically reviewed the Schwinger mechanism at zero
or finite temperature in inhomogeneous electric fields motivated by
astrophysics or terrestrial experiments. As exact solutions of the
Klein-Gordon or Dirac equation minimally coupled to inhomogeneous
electromagnetic fields are known only for a few cases, for general
electromagnetic fields, however, one has to rely on some
approximation schemes. Inhomogeneous electric fields result in a
finite size or duration effect and differs from that by a constant
field \cite{Wang-Wong,Dunne-Schubert,DWGS}. We applied the
phase-integral method to find the WKB instanton action for the mode
equations in inhomogeneous electromagnetic fields and then
calculated the pair production rate by a Sauter-type electric field
either in space or time \cite{Kim-Page02,Kim-Page06,Kim-Page07} and
an oscillating electric field. We also studied the thermal effect on
pair production by an electric field that acts for a finite period
of time \cite{Kim-Lee}. Finally, we applied the WKB instanton action
method to strange stars to calculate the electron-positron pair
production rate.

The issues not treated in  this talk are the effective action and
the back reaction of QED at zero or finite temperature. It is a
complicated task to obtain the effective action in inhomogeneous
electromagnetic fields. In canonical quantum field theory, we may
follow Ref. \cite{AHN}, according to which the effective action is
related with the scattering amplitude as
\begin{equation}
e^{i S_{\rm eff}} = e^{i \int dt d^3{\bf x} {\cal L}_{\rm eff}} =
\langle 0, {\rm out} \vert 0, {\rm in} \rangle.
\end{equation}
Thus the effective action requires a complete knowledge of evolution
of the ingoing vacuum to the outgoing vacuum, which may follow from
the vacuum wave functional from Eqs. (\ref{pa op}) and (\ref{anti
op}) for each mode. Another important issue is the QED back reaction
problem, which is described by, for instance, the scalar QED action
of the form
\begin{eqnarray}
{\cal L} = \phi^*(\partial_{\mu} + iq A_{\mu} )^2 \phi - \frac{1}{4}
F_{\mu \nu} F^{\mu \nu}.
\end{eqnarray}
The back reaction cannot be neglected for strong electric fields
because the additional electric field produced by pairs is
comparable to the applied field. In fact, positive (negative)
charges of produced pairs move in the same (opposite) direction of
the applied electric field, so the current due to pairs induces an
electric field in the opposite direction of the applied field and
overshoots it until the process is reversed, which leads to the
famous plasma oscillation \cite{KESCM,Cooper-Mottola,RVX03,RVX07}.
These issues will addressed in a future publication
\cite{Kim-Lee08}.

\acknowledgments

The author would like to thank Hyun Kyu Lee and Don N. Page  for
collaborations and useful discussions and Professor S. P. Gavrilov
and Professor Tiberiu Harko for comments on the WKB approximation
and Holger Gies and Professor Naoki Itoh for useful information. And
he also would like to thank the warm hospitality of ICRANet during
the 10th Italian-Korean Symposium on Relativistic Astrophysics,
Pescara, June 25-30, 2007. This work was supported by the Korea
Science and Engineering Foundation (KOSEF) grant funded by the Korea
government (MOST) (No. R01-2005-10404-0).

\end{document}